\documentstyle[12pt]{article}
\newlength{\extraspace}
\newlength{\extraspaces}
\newcommand{\dcc}{DCC~}

\newcommand{\be}{\begin{equation}
\addtolength{\abovedisplayskip}{\extraspaces}
\addtolength{\belowdisplayskip}{\extraspaces}
\addtolength{\abovedisplayshortskip}{\extraspace}
\addtolength{\belowdisplayshortskip}{\extraspace}}
\newcommand{\ee}{\end{equation}}

\newcommand{\bq}{\begin{eqnarray}
\addtolength{\abovedisplayskip}{\extraspaces}
\addtolength{\belowdisplayskip}{\extraspaces}
\addtolength{\abovedisplayshortskip}{\extraspace}
\addtolength{\belowdisplayshortskip}{\extraspace}}
\newcommand{\eq}{\end{eqnarray}}
\newcommand{\ra}{\rightarrow}

\begin{document}

\addtolength{\baselineskip}{.8mm}

\thispagestyle{empty}

\begin{flushright}
{\sc PUPT}-1424\\
hep-ph@xxx/y9310245 \\
 October  1993
\end{flushright}
\vspace{.3cm}

\begin{center}
{\Large Squeezed Quantum State
   of  Disoriented Chiral Condensate.} \\
\vspace{0.4in}
{\large Ian I. Kogan}
\footnote{ On  leave of absence
from ITEP,
 B.Cheremyshkinskaya 25,  Moscow, 117259,     Russia.} \\
\vspace{0.2in}
{\it  Physics Department, Princeton  University \\
 Jadwin Hall,  Princeton, NJ 08544 \\
 USA} \\
\vspace{0.7in}
{\sc  Abstract}
\end{center}

\noindent
We consider the quantum state describing the
  Disoriented Chiral Condensate (DCC) which may be
 produced in high energy collisions.  Using the
  approach suggested by  Rajagopal and Wilczek to describe
 the amplification of the long wavelength classical
  pion modes,  we consider
 the quantum-mechanical  evolution of the initial vacuum state into the
 final squeezed state describing the DCC. The  obtained wave
 function has some interesting properties which are
 discussed.


\vfill

\newpage

\renewcommand{\footnotesize}{\small}

\noindent

The possibility of the  Disoriented Chiral Condensate (DCC) production
 in high energy hadronic or heavy ion collisions   attracted a lot
 of attention in  the last years \cite{ans}-\cite{ab}.

     It is well known that QCD  Lagrangian is  invariant
 (approximately iif nonzero masses for the light $N_{f}$  quarks are
 taken into account)
 under global chiral $SU(N_{f})_{L} \times SU(N_{f})_{R}$, where
 $N_{f}$ is the number of the light flavours. This symmetry is
 spontaneously broken down to vector $SU(N_{f})_{V}$ which
 leads to $N_{f}^{2} - 1$ (quasi)goldstone bosons - pions
 (if $N_{f} = 2$) or pions, kaons and $\eta$ meson (if $N_{f} = 3$).
 The order parameter for this breaking is the vacuum expectation
 value of  quark condensate
 $< \bar{\psi}\psi>$. However one can imagine that under some special
 conditions, for example after high-energy collision, there is
 a "cool"  region  surrounded by a "hot" relatively thin expanding shell,
 which separates the internal region from the outer space. This picture
 was suggested by Bjorken and is called  now the "Baked Alaska" scenario.
   In result
 the  quark  condensate may be disoriented in isotopical space - one
 gets the DCC. After hadronization the interior disoriented
 vacuum will collapse decaying into pions. The interesting signature
 of this process will be the coherent production of either
 charged or neutral pions. There are some arguments that
 the \dcc was observed in the Centauro cosmic ray events
 \cite{centavr}.

It is
 convenient to  consider the toy  model describing
 the   chiral dynamics - the linear  sigma-model  with
  four component field $ \phi^{a} = (\sigma, \vec{\pi})$,
 where   $\sigma$ and  $\vec{\pi}$ are an
isoscalar and an isovector fields (here we use the same
 notation as in \cite{rw})
\bq
S = \int d^{4}x [ \frac{1}{2}
  \partial_{\mu}\phi^{a}\partial_{\mu}\phi^{a}
 -\frac{\lambda}{4}(\phi^{a}\phi^{a} - v^{2})^{2} + H\sigma ]
\label{action}
\eq
 where $H \sim m_{q}$  describes the small explicit chiral symmetry breaking.
 The pion mass is $m_{\pi}^{2} = H/f_{\pi} =
 \lambda(f_{\pi}^{2} - v^{2})$, where $f_{\pi} = <\sigma>$. The
 sigma mass is $m_{\sigma} = 2\lambda f_{\pi}^{2}$.
 In strong coupling limit $\lambda \ra \infty$   one gets the
   constraint  $\sigma^{2} +
 \vec{\pi}^{2} = f_{\pi}^{2}$.  In the usual vacuum one has $<\sigma> =
 f_{\pi}, ~ <\vec{\pi}> = 0$
 and since $\sigma$ is an isoscalar there is an  isoscalar
 condensate $<\bar{\psi}\psi>$  only. However one can consider
 another configuration - $<\sigma> = f_{\pi}\cos\theta$ and
 $<\vec{\pi}> = f_{\pi} \vec{n} \sin\theta$, here  $\vec{n}$ is some
 unit vector in the isospace,  which describes DCC, i.e.
 some  classical pion field configuration, which is metastable
 and  decays  after some time into pions - the signature
 for this event will be the large number of  either neutral ($\pi^{0}$)
 or charged ($\pi^{\pm}$) pions. Using the classical picture
 of \dcc which predicts equal probability for all isotopical orientation
 of  condensate one can get   \cite{ans} - \cite{kt}  the probability
 $1/\sqrt{f}$ for the fraction of neutral pion
$f = N_{\pi^{0}}/(N_{pi^{+}}+N_{\pi^{-}}+N_{\pi^{0}})$.

The question we would like to discuss here is connected with  the
 quantum description of the \dcc.  The simplest possibility
 is  to describe it by the usual coherent state.
  However the usual coherent state wave function  was critisised  in
 \cite{kt} because such a description leads to  creation of pion
 system with arbitrary large charge  fluctuations. Instead it was
 suggested in \cite{kt} that the  quantum state  must be isosinglet.
 For the state with  $2n$ neutral  pions and  total number of pions $2N$
 it was suggested:
\bq
|\Psi> \sim (2 a_{+}^{\dagger}a_{-}^{\dagger} - (a_{0}^{\dagger})^{2})^{N}|0>
\eq
Then the probability to have 2n neutral pions is \cite{hs}, \cite{kt}
\bq
P(n,N) = \frac{(N!)^{2} 2^{2N}}{(2N+1)!}\frac{(2n)!}{(n!2^{n})^{2}}
 \sim \sqrt{N/n},~~n,N >>1
\eq
and corresponds to $1/\sqrt{f}$ distribution in classical picture.

However let us note that the same distribution will be obtained
and in the case of the arbitrary  relative factor between charged
 and neutral creation operators
 $[2 a_{+}^{\dagger}a_{-}^{\dagger} - \exp(i\theta)(a_{0}^{\dagger})^{2}]^{N}$
 and one can see that  from this point of view the  zero isospin
 condition  is not
 so important - what was important indeed is that the wave
 function was constructed by the operators quadratic in
$ a^{\dagger}$ and by the construction  there are equal numbers of $\pi^{+}$
  and  $\pi^{-}$.

It is interesting to understand  what are the most natural
 class of these functions and what are the dynamical  mechanisms
 leading to the   generation of these functions. As we shall
  demonstrate here the quantum state of the  \dcc is the squeezed
   state. These  quantum states known for a long time
 in quantum optics and measurement theory ( for a review
 of squeezed states see, for example \cite{yuen} -\cite{sqrev2}).
 The simplest one-mode squeezed state is parametrized by the
 two parameters $r$ and $\phi$ and can be obtained by  acting
 on the vacuum by the squeezed operators $S(r,\phi)$
\bq
S(r,\phi)|0> = \exp[\frac{r}{2}(e^{-2i\phi} a^{2} -
 e^{ 2i\phi} a^{\dagger  2})] |0>
\label{s}
\eq
These states are minimum uncertainty states with
 $\Delta X_{1}\Delta X_{2} = 1/4$, where
 $ a(a^{\dagger}) = X_{1}\pm X_{2}$,  as well as coherent states
 $ \exp(\alpha a^{\dagger})|0>$.  However
   coherent states  have some minimal quantum noise
  and $\Delta X_{1} = \Delta X_{2} = 1/2$.
  For  the squeezed states  one can reduce the
 quantum noise for one variable (increasing it for the conjugate one)
 and $\Delta Y_{1} = e^{-r}/2, \Delta Y_{2} = e^{r}/2$ where
 $Y_{1} + i Y_{2} = (X_{1} + i X_{2}) e^{-i\phi}$. The mean particle
 number $<N> = \sinh^{2} r $.

To   get the squeezed quantum state for the \dcc let us consider
 the mechanism of the amplification of the
 long wavelength pion modes, suggested by Rajagopal and
 Wilczek in paper \cite{rw}, where the  the dynamics of
 the $O(4)$  linear sigma model after quenching was considered.
  The amplification of the long wavelength pion modes was found
  in the period immediately after quenching. This  amplification
 leads to the coherent pion oscillations, i.e. to the creation
 of the \dcc. Such a behaviour can be understood if one
 considers the equation of motion for the pion field \cite{rw}B
\bq
\frac{\partial^{2}}{\partial t^{2}}\vec{\pi}(\vec{k},t) +
[k^{2} + \lambda(<\phi^{2}>(t) - v^{2})]\vec{\pi}(\vec{k},t) = 0
\label{rw}
\eq
where we substituted the $\phi^{a}\phi^{a}$  in the  nonlinear term
 in (\ref{rw}) by its spatial average $<\phi^{2}>(t)$ - this is
 nothing but
 the Hartree-Fock or mean field approximation.
  In the initial conditions one has $<\phi^{2}> < v^{2}$ and the
 long wavelength modes of the pion field with
 $k^{2} <  \lambda(v^{2} - <\phi^{2}>)$ start growing exponentially.
 The $<\phi^{2}>(t)$  starts to oscillate near the  vacuum expectation
 value $<\sigma>$ and after some time the oscillations will be damped
 enough so that all the modes will be stable. Thus we
 see that at classical level each  long wavelength mode
  is described  by the equation for a parametrically excited oscillator
 and one  get the \dcc in result of amplification of the
 zero-point quantum fluctuations of the pion field.

  This picture is  similar to one which was discussed  in \cite{gs},
   where
    the relic gravitons production  from zero-point
 quantum fluctuations during the cosmological expansion was considered.
    For graviton  mode with momentum $ n $ the equation
$y''+[ n^{2} - (R''/R)] y = 0$
was obtained, where$ R(\eta) $ is the scale factor of the
 metric $ds^{2} = R^{2}(\eta)( d\eta^{2} - d\vec{x}^{2})$ and
a prime represents $d/d\eta$. One can see  that
  this equation  is equivalent
 to pion equation (\ref{rw}) if the scale factor $R$ is connected
 with $<\phi^{2}>(t)$ as $\lambda(v^{2} - <\phi^{2}>(t)) = R''/R$.

Our problem now  is to present the quantum mechanical  formulation
   in  terms of
  pion creation and annihilation operators and to  get the
  wave function of the \dcc.
 In the  mean field   approximation the wave function
 $|\Psi> = \prod_{i, \vec{k}} |\psi>_{i, \vec{k}}$ is the
 product of the wave functions for each mode with momentum $\vec{k}$
 and isotopical index $i$. Later we shall omit  $i$.
 The equation of motion (\ref{rw})
 for each mode $\pi(\vec{k},t)$  can be derived from the lagrangian
\bq
L_{k} = \frac{1}{2}\dot{\pi}^{2}(\vec{k},t) -
 \frac{1}{2}\Omega^{2}(\vec{k},t)\pi^{2}(\vec{k},t)
\nonumber \\
 \Omega^{2}(\vec{k},t) = \vec{k}^{2} + \lambda
(<\phi^{2}>(t) - v^{2})
\eq
The wave function $ |\psi>_{ \vec{k}}$ obeys Schr\"{o}dinger equation
\bq
i\frac{\partial}{\partial t}|\psi>_{ \vec{k}}~ =~
 H_{k}(t)|\psi>_{\vec{k}} ~ =~
[ \frac{1}{2} {\cal P}_{\pi}^{2} +
 \frac{1}{2}\Omega^{2}(\vec{k},t)\pi^{2}(\vec{k})]|\psi>_{\vec{k}}
\label{schr}
\eq
where  $\pi(\vec{k})$ and ${\cal P}_{\pi} = -i d/d \pi(\vec{k})$
 are the quantum-mechanical coordinate and momentum for the
 mode with the spatial momentum $\vec{k}$.
 One can  rewrite the Hamiltonian in (\ref{schr}) in terms of
 creation and annihilation operators which make it diagonal at
 any given moment. It is evident that we  are interested to
  get   the wave function in terms of  creation and annihilation operators
 of  ordinary pions, so  we  must  diagonalise the Hamiltonian at
 $t \ra \infty$, when the oscillation of the $<\phi^{2}>(t)$
 will be damped. Thus we define
\bq
a(\vec{k}) = \frac{{\cal P}_{\pi}  + i \omega(\vec{k})
 \pi (\vec{k})}{\sqrt{2\omega(\vec{k})}},
 ~~~~~~~
a^{\dagger}(\vec{k})  = \frac{{\cal P}_{\pi}  - i \omega(\vec{k})
 \pi(\vec{k})}
{\sqrt{2\omega(\vec{k})}}
 \eq
 where $\omega(\vec{k})  = \Omega(\vec{k},\infty) = \sqrt{\vec{k}^{2} +
 m_{\pi}^{2}}$.
It is easy to see that the Hamiltonian is
\bq
 H_{k} = \frac{1}{2}\omega(\vec{k})\bigl[1+ \frac{\Omega^{2}(\vec{k},t)}
{\omega^{2}(\vec{k})}
\bigr]
a^{\dagger}(\vec{k})a(\vec{k}) +
\frac{\omega^{2}(\vec{k})-\Omega^{2}(\vec{k},t)}{4\omega(\vec{k})}
\bigl[a^{2}(\vec{k}) + a^{\dagger 2}(\vec{k})\bigr]
\eq
The $a^{2}(\vec{k})$ and $a^{\dagger 2}(\vec{k})$
  terms in Hamiltonian is the reason
 that $H$ transforms the initial vacuum state $|0>$  into a
 squeezed state $S(r,\phi)|0>$   (\ref{s}).

To calculate the squeezing and phase  parameters $r$ and $\phi$  let us
 consider the solution of the Schr\"{o}dinger equation
 (\ref{schr}) in the coordinate (here it is $\pi$) representation
( we shall omit the label $\vec{k}$ for moment)
\bq
\psi(\pi, t) = C(t) \exp (- B(t) \pi^{2})
\label{wavefunction}
\eq
and for $B = \omega/2$ this wave function describes the vacuum
 state. For all other values this wave function describes
 the squeezed state (\ref{s}) where the parameters $r$ and $\phi$
 are connected with $B$   by the relation \cite{yuen} (see
 also  \cite{sqrev1} and \cite{gs})
\bq
 B = \frac{\omega}{2} \frac{\cosh r + \exp(2i\phi)\sinh r}{
\cosh r - \exp(2i\phi)\sinh r} \nonumber \\
 \cosh 2r = \frac{ \omega^{2} + 4 |B|^{2}}{4\omega ReB}; ~~~
\sin 2\phi = \frac{1}{\sinh 2r} \frac{Im B}{ Re B}
\label{B}
\eq
 Substituting (\ref{wavefunction}) into (\ref{schr})  one gets
 the equation for $B(t)$
\bq
i\dot{B} = 2B^{2} - \frac{1}{2}\Omega^{2}(t)
\eq
which means that $B(t)$ is related to the solution
 of the classical equation (\ref{rw})[B
\bq
B(t) = - \frac{i}{2} \frac{\dot{\psi}(t)}{\psi(t)}, ~~~~~~~~~~~~~~
\ddot{\psi}(t) + \Omega^{2}(t) \psi(t)  = 0
\eq
The last equation can be viewed as a Schr\"{o}dinger equation
 describing the wave function $\psi(t)$  of  a  "particle"
 with mass $m = 1/2$
 on a line $t$ having energy $k^{2}$ and moving through the
 potential barrier $V(t) = - \lambda
(<\phi^{2}>(t) - v^{2})$
\bq
-\frac{d^{2}\psi(t)}{dt^{2}} + \lambda (v^{2} - <\phi^{2}>(t))
 \psi(t) =
 k^{2} \psi(t)
\eq
 Far from the barrier, i.e. at $t \ra \pm \infty$ one
 has $V(\pm \infty) = - m_{\pi}^{2}$ and general solution
 of the Schr\"{o}dinger equation at $ t \ra \pm \infty$
 is the superposition of the left   and right moving waves
\bq
\psi(t) = A e^{-i\omega(k) t}  + B^{*} e^{+i\omega(k) t}
{}~~~~~~~~ t \ra +\infty
 \nonumber \\
\psi(t) = C e^{-i\omega(k) t}  + D^{*} e^{+i\omega(k) t}
{}~~~~~~~~ t \ra  -\infty
\label{abcd}
\eq
where $\omega(k) = \sqrt{k^{2} + m^{2}}$.
   Due to the unitarity the  total fluxes at $t \ra \pm \infty$
 must be equal  $|A|^{2}-|B|^{2}=|C|^{2}-|D|^{2}$ and one
 can find
\bq
A & = &\cosh r C - e^{2i\theta} \sinh r D^{*} \nonumber \\
B^{*} &  = & - e^{-2i\theta} \sinh r C +   \cosh r D^{*}
\label{ABCD}
\eq
 where $\theta$ is the scattering phase and  factor $r$
 is defined by the  probability of the transition through
 the barrier.

 Let us remember that we are starting
  from the vacuum at $t \ra -\infty$, i.e. from $B = \omega/2$,
 so one must have $C = 0$. This means that  at the left
 (large negative $t$) we have only left moving outcoming wave
  $D^{*} e^{+i\omega(k) t}$.  At the
 right (large positive $t$) one has both left and right moving waves,
  i.e.
    the incoming  $B^{*} e^{+i\omega(k) t}$
 and reflected $A e^{-i\omega(k) t}$ waves.  The tunneling
 coefficient can be obtained from  (\ref{ABCD}) by puting $C=0$
\bq[B
\frac{|D|^{2}}{|B|^{2}} = \frac{1}{\cosh^{2} r}
\eq

   Now let us calculate
  $B(t) = -(i/2) (\dot{\psi}/\psi)$
  at large positive $t$.  Using (\ref{abcd})  one  can find
 after simple calculations:
\bq
 B = \frac{\omega}{2} \frac{\cosh r + \exp[2i(\theta -\omega(k) t)]
\sinh r}{\cosh r - \exp[2i(\theta - \omega(k) t)]\sinh r}
\eq
 in complete agreement  with (\ref{B}), where now the phase factor
 $\phi = \theta - \omega(k)t$.

The squeezing parameter $r$ depends on the absolute
 value $k$  of the  mode momentum $\vec{k}$ of course
 and as we  had found it is determined by the  probability
 of tunneling through the  potential barrier
 $V(t) = \lambda
(<\phi^{2}>(t) - v^{2})$ .
Let us note that tunneling takes place precisely when
$k^{2} - V(t) <0$, i.e. when the classical long
 wavelength modes  are exponentially amplified.
  and we see once more that  squeezing is ultimately
 connected with the exponential growth of the
 classical long wavelength modes. Using the quasiclassical
 approximation it is easy to calculate the squeezing parameter
 $r(k)$:
\bq
\frac{1}{\cosh^{2} r(k)} =\exp\bigl(-2 Re \int dt \sqrt{V(t) - k^{2}}\bigr)
\eq
Thus for small $k$ and  large $r(k)$ one has
\bq
r(k) = 2 Re  \int dt \sqrt{\lambda
(v^{2} - <\phi^{2}>(t))  - k^{2}}
\eq
and the squuezing for each mode $k$  is determined by the
  function $<\phi^{2}>(t)$ - this is the only input information
 we must know to calculate the \dcc wave function.
The average number of particles in each mode is
\bq
<N_{k}> = \sinh^{2} r_{k} =
\exp\bigl(2 Re \int dt \sqrt{V(t) - k^{2}}\bigr) - 1
\eq
and we see that $<N_{k}>$ sharply decrease with the increase of $k$.
 Let us consider the simple model for $<\phi^{2}>(t)$ assuming
 that $<\phi^{2}>(t) = 0$ for $t \in (0, m_{\pi}^{-1})$ and
 then becomes equal to its usual v.e.v. One can estimate
 \bq
r(k) = \sqrt{2}\frac{m_{\sigma}}{m_{\pi}}
\sqrt{1- 2\frac{\omega^{2}(k)}{m_{\sigma}^{2}}}
\approx 6 \sqrt{1- 2\frac{\omega^{2}(k)}{m_{\sigma}^{2}}}
\eq
Of course this is the very crude estimate, however it demonstrates
 the qualitative features of the phenomenon - the
 sharp exponential dependence on $k$ and large amplification
 dactor in $r$ which is of order of $m_{\sigma}\tau$, where
 $\tau$ is the characteristic time of damping of
$<\phi^{2}>(t)$ oscillations.

Another interesting feature is the phenomenon of bunching
 and super-Poissonian statistics \cite{walls}, \cite{sqrev2}.
One may consider the second-order correlation function
\bq
g^{2}(t) = \frac{< N(t) N(0)>}{<N>^{2}}
\eq
which gives us the relative probability to measure
 two particles in an interval $t$ and $g^{2}(0)$ measures
 the probability of simultaneous detection. For coherent
 state one has $g^{2}(t) = 1$ which means that  for
 coherent state the detection events are not statistically
 dependent. For squeezed state with large $r$ one
 gets\footnote{ For small $r$, i.e. when number of
 particles is small, this formula has no physical
 meaning}  \cite{walls}
$  g^{2}(0) = 2 + \coth^{2} r > 3$
 and $g^{2}(0)$ is increasing with increasing of $k$.
The experimental measurement of $g^{2}(0)$ for pions
 with different $k$ will be extremely interesting.

Let us note that the picture we have considered here may
 have applications not only to \dcc but to the gluon condensate too.
In a recent paper \cite{gluon}  it was  shown
that a high frequency standing wave in SU(2) gauge theory is unstable
against decay into long wavelength modes which gives the
mechanism for energy transfer from initial high momentum modes
to final states
with low momentum  excitations. The obtained picture is similar to
 the amplification of the pion soft modes and it is possible
 to obtain ther same squeezed description of the final state for the
 soft gluon modes. It is interesting that because the squeezed final state
 is quadratic in field one can get gluon condensate
 $<G_{\mu\nu}^{2}>$ .

 In conclusion we would like to stress that
 in the mean field (Hartree-Fock)
  approximation the wave function
 of \dcc is the product of the  the squeezed state
  wave functions of all three pions
\bq
|\Psi>_{DCC} = \prod_{i=1}^{3}\bigl[
\prod_{\vec{k}}S(r(\vec{k}), \phi(\vec{k})) |0>_{i}\bigr]
\eq
where the universal
 functions $r(\vec{k})$ and $\phi(\vec{k})$
  are  completely determined (in the mean field
 approximation) by the only function $ <\phi^{2}>(t)$.
 This wave function leads to some
  interesting  predictions   about $k^{2}$
 dependence of the observables  which will be
 interesting to check in experiment. It is
  also interesting to go beyond the Hartree-Fock
 approximation and get the effects of correlations
 between different pion modes.

I am grateful to A.A. Anselm, D. Gross, K. Rajagopal, S.G. Matinyan
 B. M\"{u}ller,  A.M. Polyakov, W.D. Walker and F. Wilczek
  for interesting and
 stimulating discussions. This work was supported by the National
 Science Foundation grant NSF PHY90-21984.

\end{document}